# Crack Bridging Mechanism for Glass Strengthening by Organosilane Water-Based Coatings

R. Briard, C. Heitz, E. Barthel*

Surface du Verre et Interfaces
Laboratoire Mixte CNRS / Saint-Gobain ;
Saint-Gobain Recherche
39, quai Lucien Lefranc
F 93303 Aubervilliers Cedex, France

**Abstract**

We used an epoxysilane/aminosilane coating deposited from an aqueous solution to strengthen flat glass. We studied film formation, interfacial and mechanical properties of the film. The film is highly cross-linked with a 6 GPa Young's modulus and good adhesion. Our results suggest that crack face bridging accounts for most of the 75 % reinforcement in this system.

PACS : 62.20.Mk (Fatigue, brittleness, fracture, and cracks)

*Corresponding author. Tel.: +33 1 48 39 55 57; Fax: +33 1 48 39 55 62
E-mail address: etienne.barthel@saint-gobain.com

# 1. Introduction

It is common knowledge that the actual strength of glass is always lower than its intrinsic strength owing to the presence of surface microcracks, the so-called Griffith flaws [1,2]. Such defects occur at the glass surface when contact is made with other objects during fabrication or service.

The final rupture depends upon the strength and the geometry of the flaw. Stress concentration appears at the crack tip and is characterized by the stress intensity factor K. For a sharp crack of length c under an applied tensile stress σ, K is given by :

$$K = Y\sigma\sqrt{c} \qquad (1)$$

where Y is a geometric factor which depends on the shape of the sample. Failure occurs when the applied stress and flaw size give rise to a stress intensity factor which exceeds the critical stress intensity factor Kc. Kc is an intrinsic parameter of the material : Kc = 0.72-0.86 MPa.m$^{1/2}$ for soda-lime-silica plate glass.

Various techniques have been developed to strengthen glass : etching which blunts the crack tip and reduces the stress singularity, thermal tempering [3-5] and ion exchange [6] which induce surface compression and so reduce the stress "felt" by the crack. A third type of strengthening which is cheaper and easier to apply is obtained with coatings : sol-gel derived coatings [7-9] and polymeric coatings [10-14] have received attention. Hypotheses generally presented in these works to explain the reinforcement in addition to tip blunting and surface compression are energy dissipation at the crack tip for polymers, protection against water assisted corrosion and crack healing.

In this paper an hybrid "organic/inorganic" coating based on organosilanes deposited from an aqueous solution with more than 98% weight of water is studied. A polymeric network is created via epoxy/amine reaction and silanol cross-condensation while film adhesion is ensured by silane grafting on glass. Mechanical strengthening observed by a flexion test of preindented samples is correlated to the physico-chemical characteristics of the solution and of the film. The results suggests that direct bridging of the crack surfaces accounts for most of the reinforcement in this system.

**2. Experimental**

*2.1 Materials*

We use a mixture of an epoxysilane (γ-glycidoxypropylmethyldiethoxysilane) and an aminosilane (3-aminopropyltriethoxysilane) (**Table 1**).

The substrates are 2.9 mm thick soda-lime-silicate float glass and cut in (100*30) mm plates.

*2.2 Film formation*

The typical film is deposited by dipping the substrates in an aqueous silane solution (1.0% wt epoxysilane + 0.3% wt aminosilane) at a withdrawal rate of 50 cm.min$^{-1}$. The samples are dried from 10 to 30 min in air. This variation in drying time results from the preparation of large series of samples. The samples are then cured at 200°C for 20 min. Scanning electron microscopy on cross sections indicate that the film is about 100 nm thick.

Other drying times and thermal treatments were also used in specific cases and are mentioned were appropriate.

*2.3 IR studies*

Hydrolysis and condensation rates are followed by Fourier Transform Infrared (FTIR) spectroscopy using a middle IR source, a KBr beamsplitter and a MCT detector. Spectra are taken at a resolution of 4 cm$^{-1}$ for 128 scans in the range 4000 to 800 cm$^{-1}$.

The extent of reaction between amino groups and epoxy rings and silanol condensation during film drying and after thermal treatment are monitored by Reflection Absorption Infrared (RAIR). The RAIR spectra were obtained by using the same spectrometer at an incidence angle of 80° and a parallel polarization. Spectra are taken at a resolution of 4 cm$^{-1}$ for 128 scans in the range 4000 to 800 cm$^{-1}$. For RAIR, a metallic substrate is necessary [15-17], so microscope slides coated by a gold film about 1 micron thick are used.

*2.4 Film properties*

*2.4.1 Adhesion and identification of the interface*

The double cantilever beam (DCB) test is used to determine the adhesion or the cohesion of the film. A joint made of a symmetrical sandwich "glass/film/glue/film/glass" is opened by a wedge and the opening δ and the crack length L are measured. The model of Kanninen [18] gives the toughness of the joint :

$$G = \frac{3Eh^3}{16} \frac{\delta^2}{(L+0.64h)^4} \qquad (2)$$

where h and E are respectively the substrate thickness and the substrate Young's modulus.

The interface of fracture is identified using X-ray Photoelectron Spectroscopy (XPS).

*2.4.2 Elastic modulus*

The Young's modulus of the coating is determined by nanoindentation. We used a nanoindenter XP from MTS in Continuous Stiffness Measurement (CSM) mode with a Berkovich tip. The data are analysed with the Oliver and Pharr method [19].

*2.5 Tensile strength*

A microindenter is used to create controlled defects. A load of 10N is held constant for 5 seconds to create Vickers indents. Such a procedure produces a well characterised and reproducible system of cracks known as the median/radial and lateral cracks [20-23]. The indenter is oriented such that the diagonals of the indent are parallel to the edges of the specimen. Following indentation, complete removal of residual stresses is obtained by annealing the samples at 560°C for 1 hour and by cooling them very slowly (<1°.min$^{-1}$). Semi-elliptical cracks produced by this method have a surface length of 165 ± 10 μm and a depth of 55 ± 5 μm (**Figure 1**).

Failure strengths of coated and uncoated samples are measured using four point bending on a Zwick testing machine. The indented side is in tension and an adhesive tape is placed on the other side of the samples for retention of broken fragments after failure. The loading rate is 10 mm.min$^{-1}$.

For our semi-elliptical crack, Y=1.47 in equation (1) and c is the crack depth [24].

**3. Results**

*3.1 Solution study*

The hydrolysis of the silanes follows the reaction path :

$$\equiv Si\text{-}OEt + H_2O \leftrightarrow \equiv Si\text{-}OH + EtOH \qquad (3)$$

where Et stands for an ethyl group.

We studied the hydrolysis of each reagent separately by monitoring the growth of the 1045 cm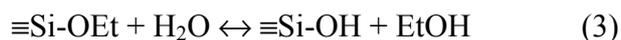$^{-1}$ band which is characteristic of the ethanol released in equation (1).

The hydrolysis of the aminosilane alone at 0.3% wt is very quick : 90% of the silane is hydrolysed in 5 min. Indeed, the polarized bond N-H permits good solubility in water which favours the hydrolysis of the ethyl groups. Fitting this hydrolysis by a kinetics of the first order, $f(t)=1-\exp(-t/\tau)$, gives a time constant $\tau \approx 1.6$ min (**Figure 2**).

It should be noted that the aminosilane sets the pH of the mixture at 10.3. Therefore, the hydrolysis of the epoxysilane at 1.0% wt was studied at two different pH. Although a first order kinetics does not match the data perfectly at pH=10.3, the 8.0 min time constant at pH=10.3 is significantly larger than at the natural pH=4.4 (2.8 min). Such a pH dependence is well known [25].

From this observation, we prepare the mixture in the following way: the epoxysilane is first prehydrolysed at its natural pH during 5 min, when 80% of the hydrolysis is completed, then the aminosilane is added.

The typical films are dip-coated 25 min after the addition of the aminosilane, when the hydrolysis of the mixture is complete according to our measurements.

*3.2 Film formation*

In our system, two kinds of reactions are expected to create a "polymeric" network : the silanol cross-condensation and the organic epoxy/amine reaction. These two reactions may occur either during drying or during the thermal treatment. They were monitored by RAIR.

*3.2.1 Spectroscopic features*

The silanol cross-condensation follows the reaction path :

$$\equiv\text{Si-OH} + \text{HO-Si}\equiv \leftrightarrow \equiv\text{Si-O-Si}\equiv + H_2O \qquad (4)$$

It is monitored by the SiOH vibration around 910 cm$^{-1}$ (**Figure 3**). The decrease of the peak indicates the loss of silanol groups. The fraction of unreacted silanol SiOH/SiOH$_{tot}$ upon drying was estimated according to:

$$\text{SiOH/SiOH}_{tot} = A/(A_{ep}+A_{am})$$

where A is the absorbance at 910 cm$^{-1}$ in the film containing the mixture of silanes and $A_{ep}$ and $A_{am}$ stand for the absorbance at 910 cm$^{-1}$ for the film containing only the epoxysilane (respectively the aminosilane) at very short drying time. We made the assumption that condensation was negligible at short drying time for the films containing each silane alone. This assumption is valid for the epoxysilane, which does not condense in solution at its natural pH and condenses only very slowly upon drying (results not shown). It is approximate for the aminosilane, which is already slightly condensed in solution and for which condensation upon drying has begun even at the early stage of drying.

The epoxy/amine cross-linking follows the reaction path :

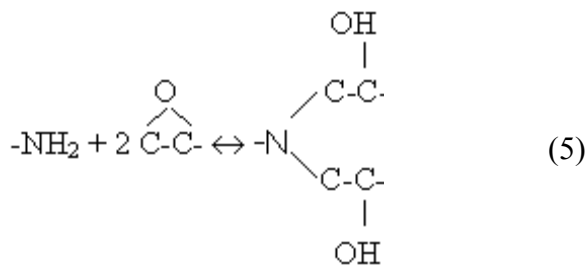   (5)

In solution, no epoxy/amine reaction was observed. Within the film, the reaction is monitored by following the epoxy bands

(3000 and 3057 cm$^{-1}$) and the amino bands (3300 and 3358 cm$^{-1}$) (**Figure 4**).

*3.2.2 Condensation and cross-linking*

Upon drying in ambient air, the total amount of silanols in the film drops to about 15% of its original value within 10 hours. However, only the final thermal treatment ensures that condensation is complete.

After five hours of drying in air, the amino groups have totally disappeared. Because the solution is formulated with an excess of epoxysilane, epoxy groups remain even after thermal treatment This complete reaction of the amino groups suggests a thorough mixing and an homogeneous distribution of the silanes, which can react together (**Figure 5**). In summary, these spectroscopic evidences show that a dense network of silanes is obtained. Although the thermal treatment is necessary to complete the condensation, room temperature drying is sufficient to polymerize most of the material.

*3.3 Film properties*

To characterize the film properties, samples dipped in a 30 min aged solution, dried for 10 to 20 min and cured at 200°C for 20 min were prepared.

*3.3.1 Adhesion and identification of the failure interface*

DCB adhesion tests on these films evidence a high joint toughness. However, the XPS shows that the fracture occurs at the glue/film interface. Calculations based on the Kanninen model give an interfacial toughness G ≈ 8 J.m$^{-2}$ (**Figure 6**). The interfacial toughness reflects the work of adhesion but is not directly equal to it because of additional plastic dissipation in the glue. In the present experimental configuration, we estimate the enhancement factor as about four [26]. Thus, we may say that the layer toughness and the layer adhesion is higher than about 2 J.m$^{-2}$. Indeed, silanes were specially chosen for their good reactivity with the glass surface silanols, consistent with the good adhesion on glass. In the same way, we have seen above that covalent bonds are created within the film in agreement with the high cohesion energy of the coating.

*3.3.2 Elastic modulus*

Nanoindentation was performed on a 100 nm thick film. The effective modulus as a function of penetration exhibits the typical "S" curve for coated substrates (**Figure 7**). At large penetration, the substrate stiffness is obtained ($E_{glass}$ ≈ 75 GPa) which signals the absence of delamination, another evidence for good adhesion. Extrapolation at low penetration returns the intrinsic film modulus around 6 GPa. Thus, the film modulus is about 10 times smaller than the substrate and somewhat larger than a typical epoxy resin (E ≈ 2 GPa).

*3.3.3 water stability*

To check the water stability of the film, a coating has been tested during 2 months by contact angle. The specimen was stored in room conditions (20°C and 40% humidity rate) and was tested periodically by contact angle. A constant 65° angle was observed which means that the film is rather hydrophobic as expected from the loss of the hydrophilic silanol and amino groups. Water durability and permeability have also been studied by transmission FTIR: the coated sample is immersed in deionised water and infrared spectra are recorded daily after taking out the sample of the water bath and blowing it dry with nitrogen gas. The OH peak (3400 cm$^{-1}$) and the CH peaks (2868 and 2931 cm$^{-1}$) are monitored (**Figure 8**). The CH and OH peak areas are almost constant, which indicates a good water durability and low water permeability of the film.

*3.4 Glass strengthening*

The rupture stresses measured by flexion tests are displayed on **Figure 9** for various coatings. In comparison with uncoated (but indented and annealed) samples, we observed an improvement of 75% of the rupture stress for a thermal treatment of 200°C during 20 min : for uncoated samples, $\sigma_r$ = 76±2 MPa while $\sigma_r$ = 133±9 Mpa for coated samples. The distributions are narrow which proves that the process is highly reproducible.

Moreover, we observed that no reinforcement is obtained when either of the two reagents is applied separately. Finally, for the silane mixture, we observed a reinforcement of 35% for samples dried 24 hours in air and a strengthening of 64% for a 20 min thermal treatment at 110°C.

## 4. Discussion

The different possible mechanisms for glass strengthening will be considered now. We note that the formation of a network is clearly necessary for reinforcement. The reagents applied separately cannot form a network. Indeed, a difunctional silane like the epoxysilane polymerises linearly. In the same way, a trifunctional silane like the aminosilane tends to form small clusters at basic pH. It is only the mixture of the two silanes which forms a cross-linked network.

*4.1 Energy dissipation by plasticity*

J.B. Denis [27] proves in his thesis work that the reinforcement is related to the yield stress of the coating. In his case, he deals with epoxy resins which present long chains able to deform plastically.

Our system is characterized by short chains and is highly cross linked : it is rather a brittle material. It seems improbable that such strengthening by plasticity may occur here.

*4.2 Thermal stresses*

This mechanism suggests that thermal compressive stresses generated by the coating in the glass surface induce the strengthening. Following Hand et al. [11], to obtain the balancing compressive stress in the glass, we assume that the stresses in both the coating (which coats both sides of the glass) and the substrate are uniformly distributed, in which case the stress in the substrate is given by :

$$\sigma = -\frac{E_{coating}}{1-\nu_{coating}} \frac{2 t_{coating}}{t_{glass}} (\alpha_{coating} - \alpha_{glass}) \Delta T \qquad (6)$$

where E, ν, α and t are the Young's modulus, Poisson's ratio, thermal expansion coefficient and thickness of the glass substrate or coating, ΔT is the difference in temperature between the curing temperature and room temperature.

The thermal expansion coefficient of the coating is unknown. Assuming $10^{-4}$ °C$^{-1}$ as an upper bound, for a thermal expansion coefficient of the glass around $10^{-5}$ °C$^{-1}$, a Young's modulus of the coating of 6 GPa, a Poisson's ratio of 0.2, ΔT=200°C and a thickness ratio of $10^{-4}$, the compressive stress in the glass is ≈ 0,01

MPa. This stresses are clearly insufficient to explain the strengthening mainly because of the negligible thickness of the film.

*4.3 Water protection*

The water corrosion of glass is a well established phenomenon [28-31]. The classical theory involves the chemical reaction of water with silica which is the reverse of equation (4).

Such behaviour is very important in static fatigue of glass when K<Kc : on soda-lime glass, the sub-critical crack growth has been described in details. It was shown that the water partial pressure has an impact on the sub-critical crack velocity but not on the Kc [31].

In our study, we deal with an hydrophobic film which can locally, at the crack tip, decrease the amount of water. However, since water has a minor effect on Kc, the increase of Kc we observe is not due to a water barrier effect, although protection against corrosion may be present.

*4.4 Crack healing*

Crack healing results from the presence of tensile stresses within the crack. These closure stresses reduce the stress intensity factor at the crack tip as

$$K_{coated} = K_{uncoated} + K_{closure} \qquad (7)$$

where the negative stress intensity factor resulting from the closure stresses is

$$K_{closure} = 2Y'\sqrt{\frac{c}{\pi}}\int_0^c \frac{\sigma(x)}{\sqrt{c^2-x^2}}dx \qquad (8)$$

Y' is a geometric factor and σ(x) the (negative) closure stress distribution.

We discuss the possible origins for these closure stresses.

*4.4.1 Thermal mismatch*

Following Hand et al. [11], we can consider a closure stress generated within the crack by thermal expansion mismatch stresses between the coating and the glass. If we consider that

$$\sigma_{closure} = -\frac{E_{coating}}{1-\nu_{coating}}(\alpha_{coating} - \alpha_{glass})\Delta T \qquad (9)$$

With our previous estimates of the thermal coefficient, the closure stresses have an order of magnitude consistent with the observed value. However, in this mechanism, the closure stresses and the closure stress intensity factor are directly proportional to the temperature difference ΔT. In this case, we would expect the reinforcement to drop by a factor of 2 for the 110 °C thermal treatment and to vanish when the film is cross-linked at room temperature. This is in contradiction with the observed values: the reinforcement is almost unaffected when the thermal treatment temperature is reduced to 110 °C, and is still significant for room temperature drying. This leads to the conclusion that thermal mismatch between the coating and the glass is not the dominant mechanism for closure stresses in the present system.

*4.4.2 Elastic healing*

Another mechanism for the generation of closure stresses is the elastic response of the material filling the crack, or crack face bridging. A simple approximate model is now developed to account for this effect.

If we assume a linear edge crack with a uniform remote loading $\sigma_0$ normal to the crack (**Figure 10**), this loading gives rise to a displacement of the crack faces [32] :

$$u \approx -\frac{K}{E^*}\sqrt{\frac{2r}{\pi}} \qquad (10)$$

with r = c-x and $E^* = \frac{E_{glass}}{1-\nu_{glass}^2}$ in plane strain.

If the crack is filled with a perfectly elastic material with an oedometric modulus E, then the displacement *u* induces a deformation of the filling materials, which in turn exerts a stress on the crack faces. To estimate this closure stress, we would need to know the crack shape, which is not possible. We therefore simply assume that the crack has a constant width h. Then, the deformation is $u/h$ and the stress distribution inside the crack is

$$\sigma(x) = \frac{E\,u(x)}{h} \qquad (11)$$

As a result the remote loading $\sigma_0$ induces a stress intensity factor K which obeys the self-consistency equation

$$K_{closure} = -\beta K_{coated}\frac{E}{E^*}\frac{c}{h} \qquad (12)$$

with $\beta = \frac{4\sqrt{2}(\sqrt{2}-1)}{\pi}Y'$

Finally, combining equations (7), (8), (10) and (11) gives

$$K_{coated} = \frac{K_{uncoated}}{1+\beta\dfrac{E}{E^*}\dfrac{c}{h}} \qquad (13)$$

Defining the crack form factor as $c/h$ and the modulus ratio as $E/E^*$, a crack with a form factor of 10 and a modulus ratio of 0.1 – which are reasonable estimates in our case – will exhibit a 60 % stress intensity factor drop. Although this simple model is approximate because equation (8) will lose its validity when the closure stresses increase and because the exact crack shape is unknown, it shows that this bridging mechanism provides a consistent explanation for the reinforcement of glass by a soft crack filling material.

## 5. Conclusions

Indented glass coated with an epoxysilane/aminosilane film of 100 nm thickness exhibits a significant increase in strength of 75%. The film, which has a 6 GPa Young's modulus, shows a good adhesion. The low water permeability of the film indicates that the film may act as a water barrier against environmental corrosion for sub-critical crack growth. However this mechanism cannot be responsible for the observed increase of the critical rupture stress. We have shown that closure stresses generated within the crack by thermal expansion mismatch stresses between the coating and the glass play a minor role in the reinforcement, and that closure stresses generated within the crack by the elastic response of the filling material are probably the dominant strengthening mechanism in the present system.

**Acknowledgments**

The authors are grateful to Dr S. Besson and S. Lohou for many helpful discussions.

**References**

[1]   A.A Griffith, Philos. Trans. R. Soc. A221 (1920) 163
[2]   A.A Griffith, Int. Congr. For Applied Mechanics, Delft. Vol. 61 (1924) 55.
[3]   L .Daudeville, H. Carre, Journal of Thermal Stresses 21 (1998) 667
[4]   C. Guillemet, J. Non-Cryst. Solids 123 (1990) 415

[5]  T.F. Soules, R.F. Buskey, S.M. Rekhson, A. Markovski, J. Am. Soc. 70 (1987) 90
[6]  A.K. Varshneya, W.C. Lacourse, Fusion and Processing of Glass, third Int. Conference (1992) 365
[7]  B.D Fabes, D.R. Uhlmann, J. Am. Ceram. Society 73 (1990) 978
[8]  B.D Fabes, G.D. Berry, J. Non-Cryst. Solids 121 (1990) 357
[9]  B.D Fabes, W.F. Poyle, W.F. Zelinski, L.A. Silverman, D.R. Uhlmann, J. Non-Cryst. Solids 82 (1986) 349
[10] J.E Ritter, M.R. Lin, Glass Technol. 32 (1991) 51
[11] R.J. Hand, B. Ellis, B.R. Whittle, F.H. Wang, J. Non-Cryst. Solids 315 (2003) 276
[12] J.G.R. Kingston, R.J. Hand, Phys. Chem. Glasses 41 (2000) 1
[13] R.J. Hand, F.H. Wang, B. Ellis, A.B. Seddon, Phys. Chem. Glasses 39 (1998) 305
[14] F.H. Wang, R.J. Hand, B. Ellis, A.B. Seddon, Phys. Chem. Glasses 36 (1995) 201
[15] R.G Greenler, The J. of Chem. Phys. 44 (1966) 310
[16] R.G Greenler, The J. of Chem. Phys. 50 (1969) 1963
[17] R.G Greenler, R.R. Rahn, J.P. Schwartz, J. of Catalysis 23 (1971) 42
[18] M.F. Kanninen, Int. Journ. Of Fracture 9 (1973) 83
[19] W.C. Oliver, G.M. Pharr, J. of Mat. Research 7 (1992) 1564
[20] B.R. Lawn, A.G. Evans, D.B. Marshall, J. Am. Ceram. Soc. 63 (1980) 574
[21] D.B. Marshall, B.R. Lawn, A.G. Evans, J. Am. Ceram. Soc. 65 (1982) 561
[22] R.F. Cook, G.M. Pharr, J. Am. Ceram. Soc. 73 (1990) 787
[23] V.M. Sglavo, D.J. Green, J. Eng. Fract. Mech. 55 (1996) 35
[24] Y. Murakami, Stress Intensity Factors Handbook Vol. 2, The Society of Materials Science, Japan, Pergamon Press 1987
[25] E.P. Plueddeman, Silane Coupling Agents, Plenum Press, New York 1982
[26] E. Barthel, O. Kerjan, P. Nael and N. Nadaud, to appear in Thin Solid Films
[27] J. B. Denis, Renforcement du verre par un revêtement organique, Doctoral Dissertation, Université Rennes I, France (2002)

[28] R. Gy, J. Non-Cryst. Solids 316 (2003) 1
[29] M. Tomozawa, Phys. Chem. Glasses 39 (1998) 65
[30] S.W. Freiman, Strength of Inorganic Glass, Plenum Press 1985
[31] S.W. Wiederhorn, J. Am. Ceram. Soc. 50 (1967) 407
[32] Fracture of Brittle Solids, B.R. Lawn, T.R. Wilshaw, Cambridge University Press 1975

**Captions**

**Figure 1**. Micrograph of a model semi-elliptical flaw produced by a 10N/5s Vickers indentation with (1) median/radial crack and (2) lateral crack. 2a = 165 ± 10 µm and c = 55 ± 5 µm.

**Figure 2.** Hydrolysis of aminosilane (0.3% wt, natural pH=10.5) and of epoxysilane (1.0% wt) for 2 different pH: natural pH (≈4) and mixture pH (≈10). Fits by $f(t)=1-\exp(-t/\tau)$

**Figure 3.** Cross-condensation studied by RAIR on a sample dipped in a 30 min aged solution, dried during different times (shown in the legends) and cured (when corresponding) at 200°C for 20 min. The decrease of the SiOH peak at 910 $cm^{-1}$ is an evidence of cross-condensation.

**Figure 4.** Organic epoxy/amine reaction studied by RAIR on a sample dipped in a 30 min aged solution, dried during different times (shown in the legends) and cured (when corresponding) at 200°C for 20 min. The decrease of the epoxy bands (3000 and 3057 $cm^{-1}$) and the total disappearance of the amino groups (3300 and 3358 $cm^{-1}$) are an evidence of the organic reaction. The excess of epoxysilane is confirmed by the presence of the epoxy band even after the thermal treatment.

**Figure 5.** Dependence with the drying time or with the thermal treatment (24 h drying and curing at 200ºC for 20 min) of the cross-condensation and of the organic epoxy/amine reaction studied by RAIR on a sample dipped in a 30 min aged solution and dried during different times.

**Figure 6.** Results of double cantilever beam test used to determine the adhesion of the film on glue in a symmetrical sandwich "glass/film/glue/film/glass" opened by a wedge : the opening δ and the crack length L are measure, h is the substrate thickness.

**Figure 7.** Elastic modulus obtained from a Berkovich nanoindentation experiment performed on a 0,1 µm thick film.

**Figure 8.** Water durability studied by transmission FTIR performed on a sample dipped in a 30 min aged solution and cured at 200°C for 20 min.

**Figure 9.** Rupture stress obtained from 4 point bending. Comparison between reagents alone and silanes mixture . Coated samples are dipped

in a 30 min aged solution and dried 24 hours in air or cured at different temperatures for 20 min.

**Figure 10.** Linear edge crack filled by a coating of Young's modulus E with a uniform remote loading $\sigma_0$ normal to the crack.

**Table:** structural formulae of : (a) the epoxysilane and (b) the aminosilane.

**Figures**

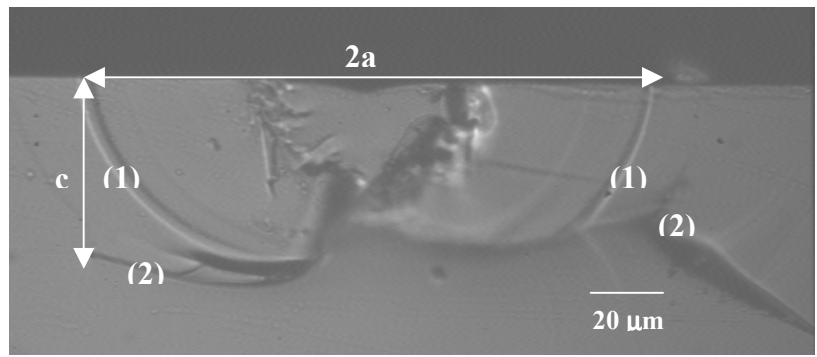

**Figure 1**

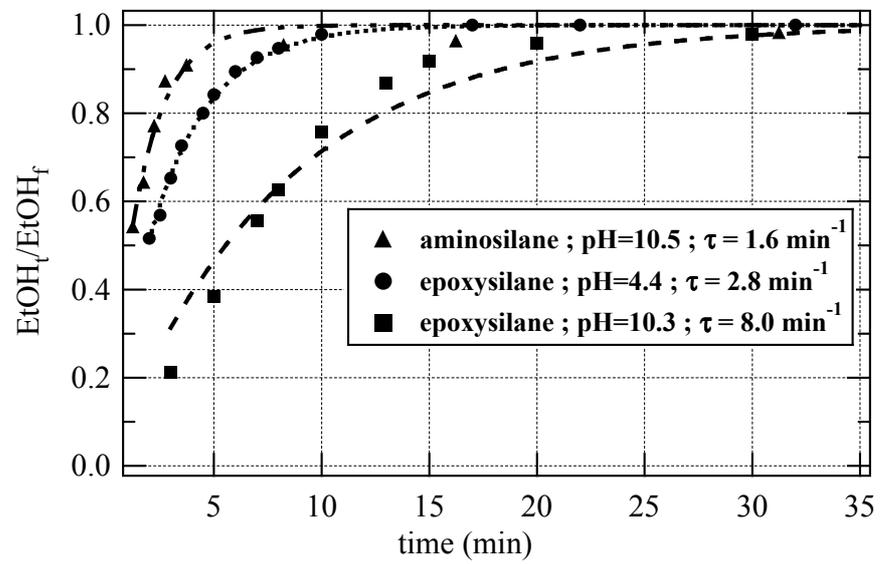

**Figure 2**

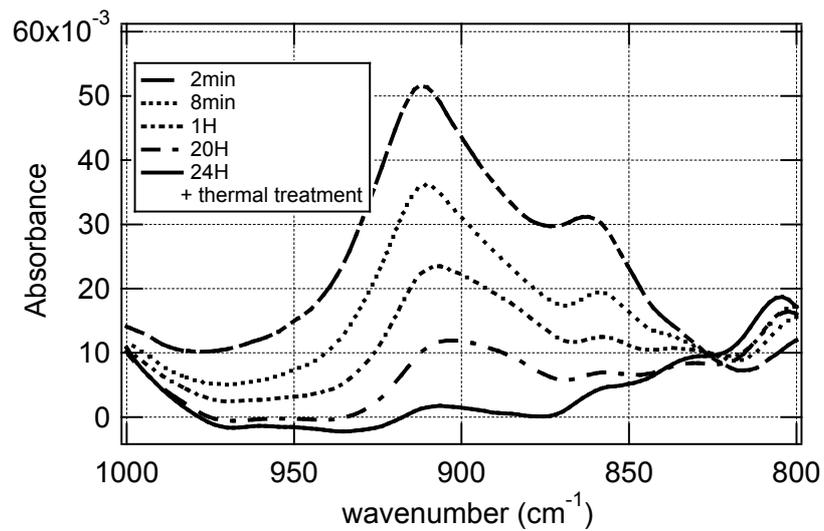

**Figure 3**

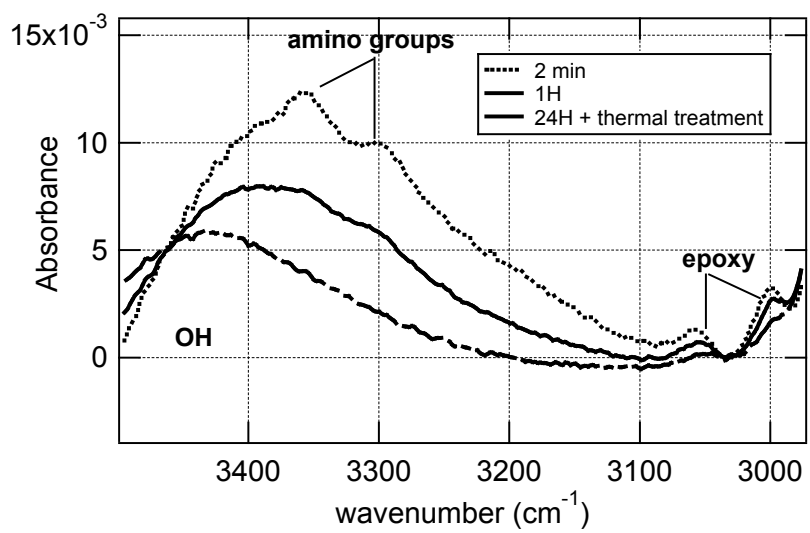

**Figure 4**

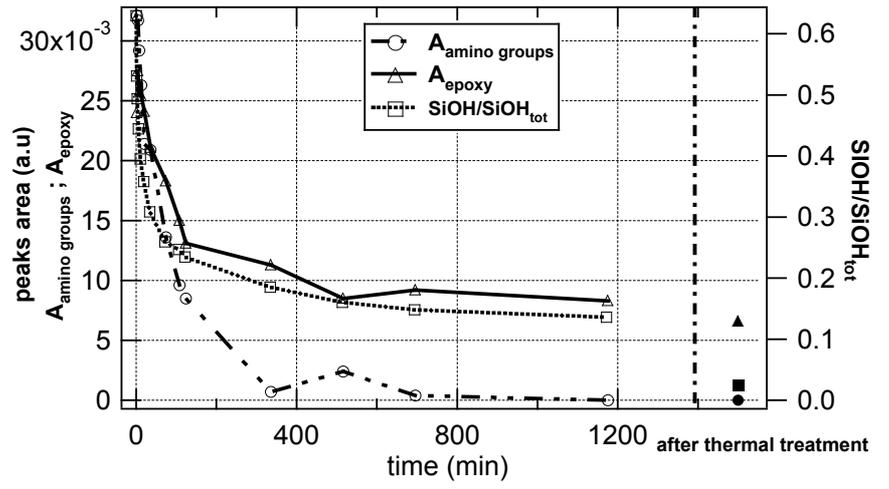

**Figure 5**

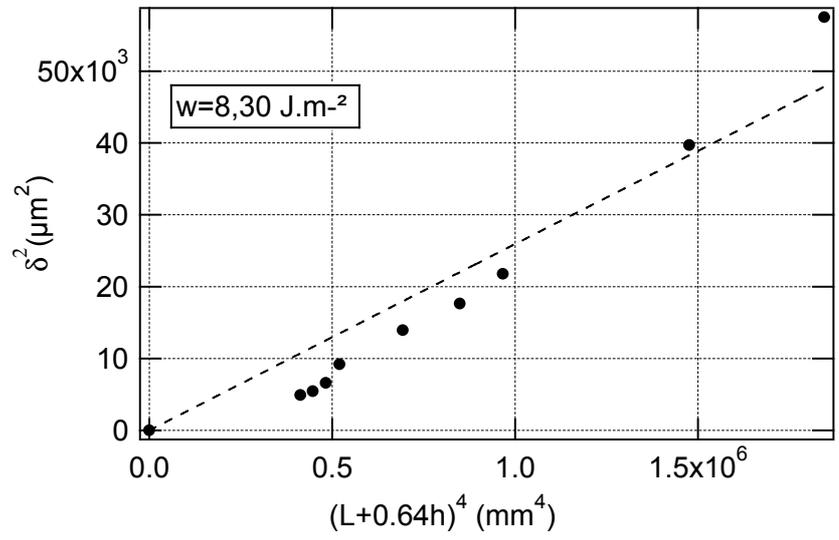

**Figure 6**

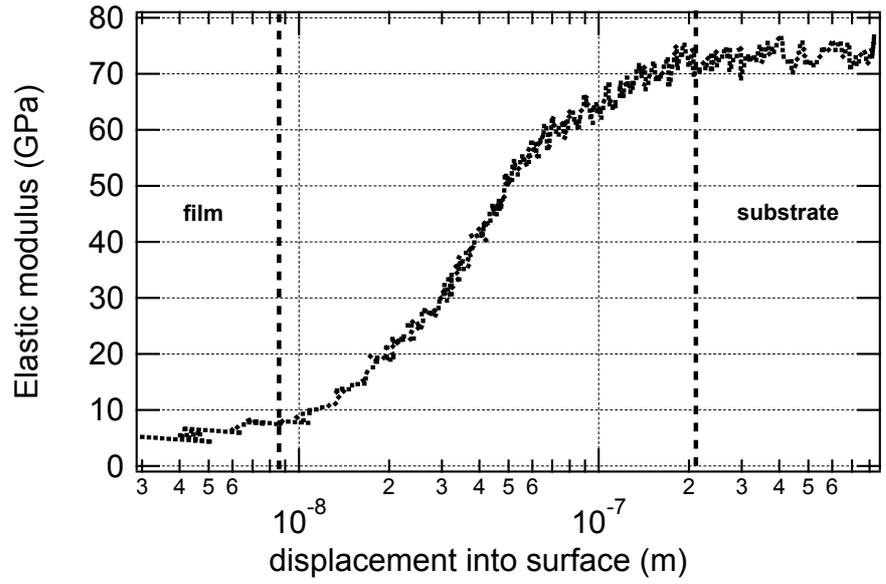

**Figure 7**

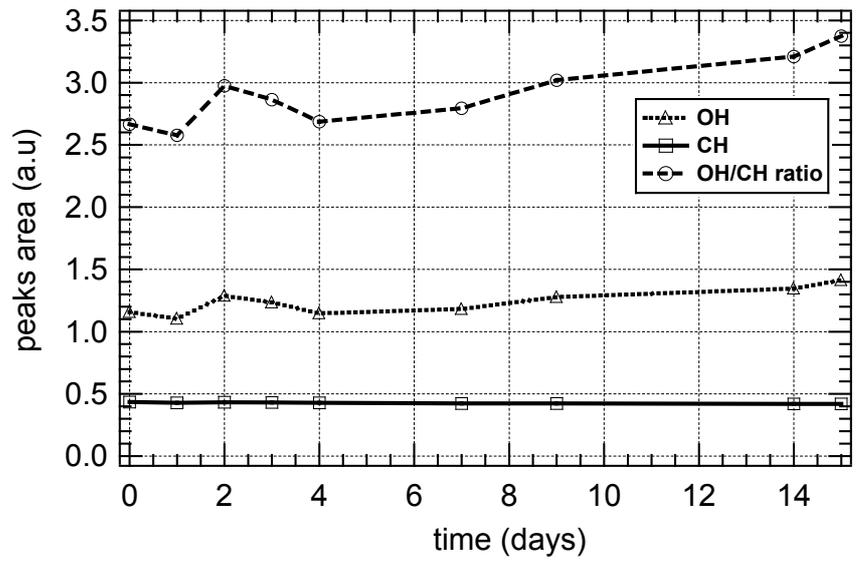

**Figure 8**

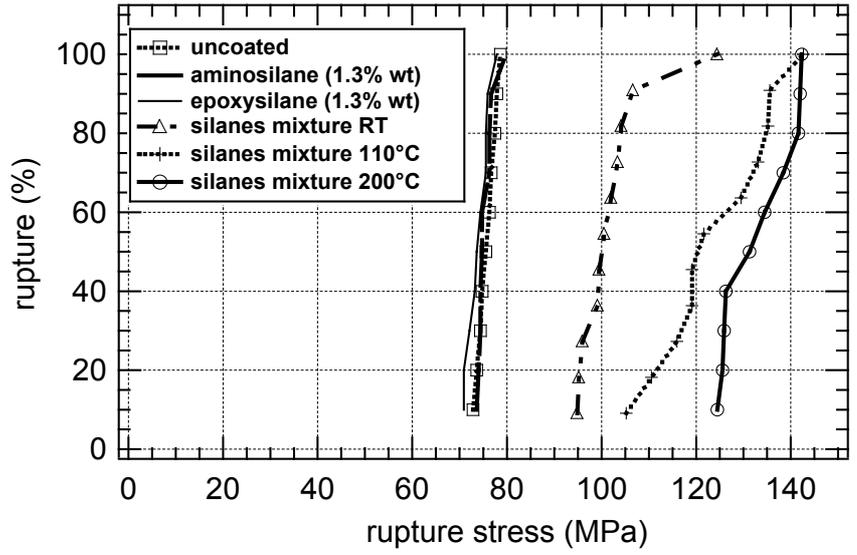

**Figure 9**

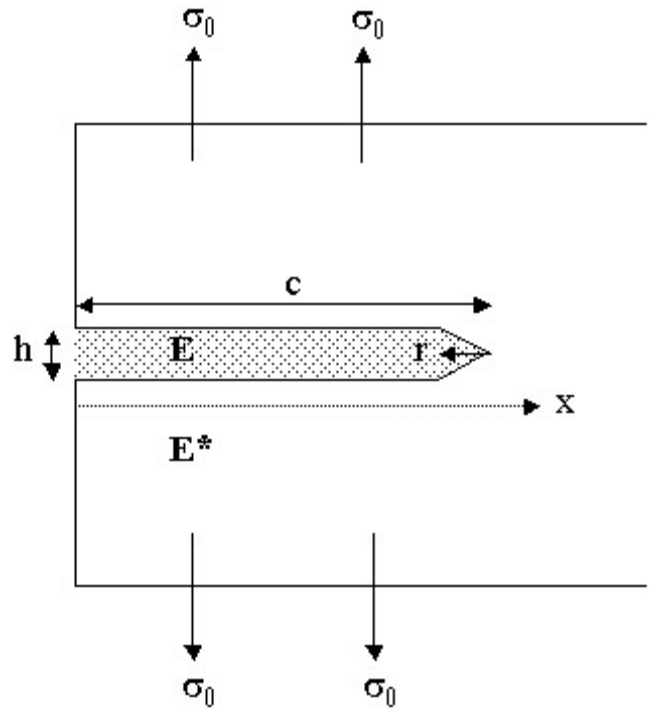

**Figure 10**

**Table**

| | |
|---|---|
| (a) | $\underset{\diagdown O \diagup}{CH_2\text{-}CH}\text{-}CH_2\text{-}O\text{-}(CH_2)_3\text{-}Si(CH_3)(OCH_2CH_3)_2$ |
| (b) | $NH_2\text{-}(CH_2)_3\text{-}Si(OCH_2CH_3)_3$ |